\pdfoutput=1

\documentclass[journal]{IEEEtran}
\ifCLASSINFOpdf
   \usepackage[pdftex]{graphicx}
\else
\fi
\begin{document}
%
\title{Simulations and Image Reconstruction for the High Resolution CaLIPSO PET Scanner for Brain and Preclinical Studies}
%
%
%

\author{Olga~Kochebina,~\IEEEmembership{SPP-IRFU-CEA,~IMIV-SHFJ-CEA,}
 S\'{e}bastien~Jan,~\IEEEmembership{IMIV-SHFJ-CEA,}
 	Simon Stute,~\IEEEmembership{IMIV-SHFJ-CEA,}
	Viatcheslav~Sharyy,~\IEEEmembership{SPP-IRFU-CEA,}	
	Patrice~Verrecchia,~\IEEEmembership{SPP-IRFU-CEA,}	
	Xavier~Mancardi,~\IEEEmembership{SPP-IRFU-CEA,}
 and~Dominique~Yvon,~\IEEEmembership{SPP-IRFU-CEA}
 
}

%
%

\markboth{Prepared for IEEE TRPMS journal}%
{Kochebina \MakeLowercase{\textit{et al.}}: Simulations and Image Reconstruction for CaLIPSO PET Scanner}
%



\maketitle

\begin{abstract}
The foreseen CaLIPSO Positron Emission Tomography~(PET) scanner is expected to yield simultaneously a fine image resolution, about 1~mm$^3$, and a high contrast. 
In this paper we present results of simulations for the full CaLIPSO PET scanner with a ``cube" geometry. We quantify by simulations the expected image resolution and Noise Equivalent Count Rates and compare them to the performance of the most efficient clinically used PET scanner, the High-Resolution Research Tomograph by Siemens. We bring up the issues of the image reconstruction for a scanner with high spatial resolution. We also present simulated brain images for [$^{18}$F]-FDG and [$^{11}$C]-PE2I tracer distributions. Results demonstrate the high potential of the CaLIPSO PET scanner for small animal and brain imaging where combination of high spatial resolution and efficiency is essential.
\end{abstract}


%
\IEEEpeerreviewmaketitle

\section{Introduction}
%
%
%
%
\IEEEPARstart{H}{igh} spatial resolution Positron Emission Tomography~(PET) is an advanced method for diagnosis and investigate mainly in neurology studies to exam neurodegenerative diseases such as Alzheimer's, Parkinson's, Huntington's diseases or multiple sclerosis. According to the radioactive tracer distribution it is possible to monitor the activity of cells and various processes such as the transport of substances, gene expression and the interaction between a ligand and a receptor.
It is also of interest in oncology for the detection of small tumors and small metastases. High image resolution imager is also useful in preclinical researches on small rodents.
Clinical Magnetic Resonance Imaging (MRI) scanners provide high image resolution but are less sensitive to monitor of biochemical cellular activity. 
Thus, high resolution PET imagers would be an excellent complement to MRI for diagnosis and medical research.

\

\begin{figure}[!t]
\centering
\includegraphics[width=3.5in]{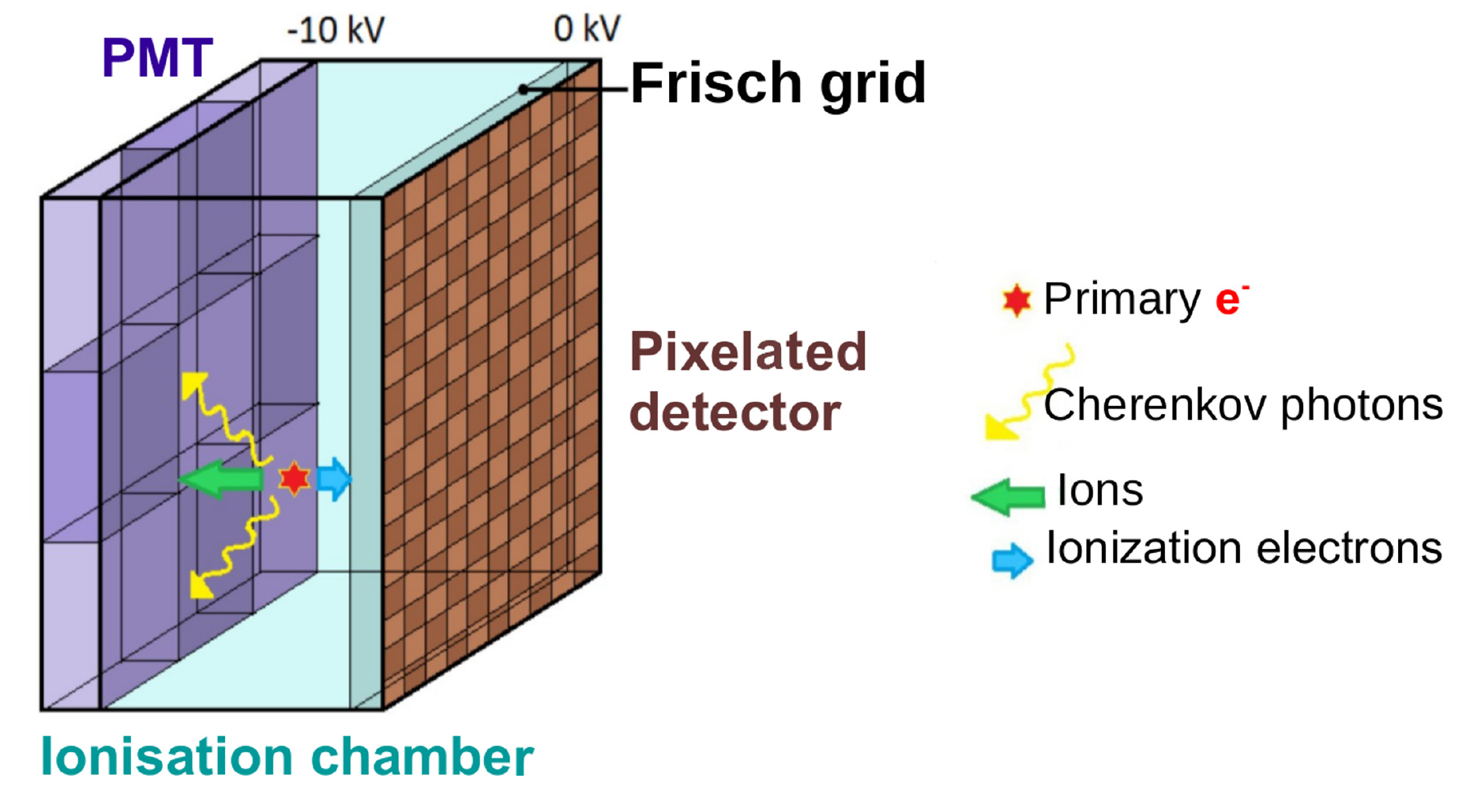}
\caption{Principle scheme of the CaLIPSO detector. The primary electron created by 511-keV $\gamma$ produces Cherenkov light and ionizes the medium. Both signals are used to measure arrival time, 3D position and energy.}
\label{fig1}
\end{figure}

The foreseen CaLIPSO\footnote{French acronym for Liquid Ionization Calorimeter, Scintillation Position Organometallic} PET scanner~\cite{CaLIPSO} is a high resolution imager for human brain and preclinical studies. It combines simultaneously a high detection efficiency, a 3D spatial resolution of 1~mm$^3$~(Full Width at Half Maximum, FWHM) and a coincidence resolving time of about 150~ps or better, allowing for a high image contrast. 
High performances of the proposed PET scanner are possible thanks to the concept of double detection illustrated in Fig~\ref{fig1}, An elementary cell of the PET imager is filled with an innovative liquid, trimethyl bismuth (TMBi)~\cite{CaLIPSO:TMBi}, It contains 82\% by weight of Bismuth, the highest-Z radioactively stable element. Consequently, TMBi allows a photoelectric conversion yield of $\sim$47\%~\cite{CaLIPSO:TMBi}, better than any other detector used or proposed for PET imaging in nuclear medicine. Photoelectric photon conversion or Compton scattering generate a so-called ``primary" electron in the liquid TMBi, which produces two types of signal. The electron emits Cherenkov light while propagating in the TMBi. However, the number of photons is quite low and light collection with efficient photomultipliers (PMT) is required. We plan to work with Micro-Channel Plate PMT (MCP-PMT) which provide also an excellent time resolution. The light detection~\cite{CaLIPSO:optic} is used for the precise measurements of the interaction time. The same primary converted electron ionizes the medium and generates free electrons which drift along a strong electric field, pass through a Frisch grid and are collected by a pixelated detector. The pixel size used to collect ionization is 1~mm$^2$ which ensures a high spatial resolution. The measurement of the ionization drift time allows to calculate the third coordinate or depth of interaction~(DOI) with 1~mm precision~\cite{CaLIPSO}, This ionization signal is used not only for the position estimate but also for energy measurements with a precision of~$\sim10\%$~(FWHM)~\cite{CaLIPSO}, 

Moreover, due to fast Cherenkov light emission and detection an excellent time resolution, coincidence resolving time $\sim$150~ps~(FWHM), is also expected. This makes possible to use time of flight technique~(TOF) to improve signal to noise ratio in final images. However, in this paper we let aside this property of the CaLIPSO scanner in the simulations as the proof of concept for the detection methods are still under validation with ionization and optical prototypes.

\

The main aim of CaLIPSO project is to develop high resolution PET scanner with reconstruction of the DOI. This problematic has been the subject of many developments during last decades. The main attempts we present in the following for the illustration of the effort that is already made.

In order to increase the spatial resolution and obtain DOI PET it was proposed to pixelaze scintillation crystals. For example, in~\cite{Vandenbroucke,Yamaya,Yoshida} the authors suggested to use array modules built out of scintillation crystal of the size of 1~mm$^3$ or less. The foreseen submillimetric 3D resolution was illustrated on prototype measurements and Monte Carlo simulations. However, the main limitation of this method is a drop of detection efficiency with crystal size.

To overcome this limitation, monolithic scintillator crystals, LaBr3:Ce, coupled with digital silicon photomultipliers was proposed. In the first experimental characterization of such a TOF PET detector 1~mm spatial resolution~\cite{Seifert} and coincidence resolving time of $\sim$160-180~ps (FWHM)~\cite{vanDam} were demonstrated. 

Completely different technology was also proposed to gain in the spatial resolution: CdTe semiconductor detectors. It has been shown that a PET scanner based on these detectors can have in transaxial plane 2.3~mm--4.8~mm resolution~\cite{Shiga}, Time resolution of such machine is low, 6.8~ns~(FWHM), and TOF cannot be used. Thus, the image contrast is enhanced with another technique: a high energy resolution, 4.1\% for 511~keV $\gamma$, is used to reduce the scatter coincidences background.

Another idea how to enhance spatial resolution without drop of detection efficiency is to use a liquid instead of crystals. For example, a PET scanner based on a liquid Xenon detectors ~\cite{Xenon} was an inspiration for CaLIPSO. The double signal detection is also used: scintillation light and ionisation signals. Studies on detector prototypes showed a submillimeter 3D spatial resolution and energy resolution better than 10\%~(FWHM).

For the high spatial resolution scanner, it is important to have low statistical noise in each voxel. One of the possibility to enhance the signal-to-noise ratio of the image is to use the TOF technique. In~\cite{Korpar} authors propose to use the Cherenkov radiation instead of scintillation light as in this case the time resolution is not limited by the scintillation decay time. Using PbF$_{2}$ crystal coupled to MCP-PMT they demonstrated the time resolution of about 70 ps (FWHM) for a single detector~\cite{Korpar}, The drawback of this approach is a low number of the optical photons produced in the Cherenkov radiator, and thus, low intrinsic detection efficiency ($<$10\%). 
The CaLIPSO detector also uses the Cherenkov radiation as explained above. The first prototypes demonstrated
the possibility to reach a much higher efficiency of the detection, of the order of 35\%, by using the optimized light collection in the TMBi liquid~\cite{Ramos},

Although new proposed technologies and methods are promising, an existing benchmark scanner, the HRRT (High Resolution Research Tomograph) by Siemens~\cite{HRRT}, is already used for clinical applications in $\sim$~20 PET centres around the world. It has a spatial resolution of approximately 2.3~mm to 3.2~mm (FWHM) in the transaxial direction. This machine without TOF uses two layers of scintillation crystals, LSO and LYSO, connected to photomultiplier tubes. This scanner combines high image spatial resolution with high sensitivity~\cite{HRRT}, In this paper we use this advanced clinical scanner for a comparison with the simulated CaLIPSO PET system. 
 
We previously simulated the CaLIPSO imager with a HRRT-like geometry~\cite{Kochebina}, In this paper, we introduce a geometry for the scanner with a larger solid angle. Through the use of Monte Carlo simulations, we measure its sensitivity and transaxial spatial resolution and perform a preliminary 2D image quality evaluation. It is important to mention that the current status of the actual detector cell design, which is still ongoing, did not justify the costly effort for developing a complete 3D reconstruction software taking TOF into account. This is further discussed throughout the paper.

In ``Materials and Methods" section we present the simulation approaches and the evaluation setups. 
``Results" section gives the Noise Equivalent Count Rates calculation, estimations for image resolution using capillary sources, reconstructed images of the Derenzo phantom and simulated brain images for two tracers, [$^{11}$C]-PE2I and [$^{18}$F]-FDG. In ``Discussion" and ``Conclusion" sections we give overview of the results and the perspectives of the CaLIPSO project.

\section{Materials and Methods}

\begin{figure*}[!t]
\centering
\includegraphics[width=\textwidth]{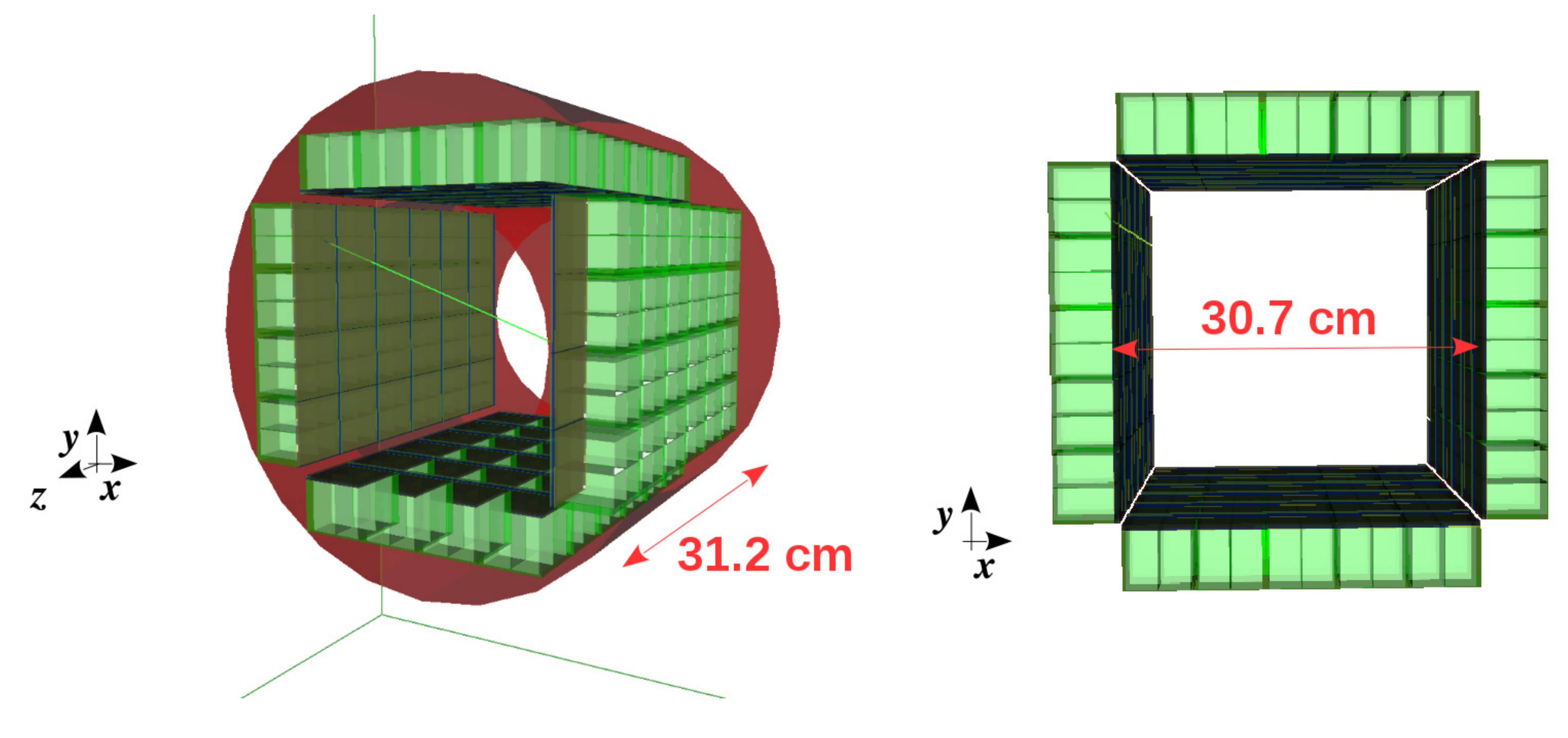}
\caption{Full brain scanner used in the simulation for CaLIPSO PET.}
\label{fig2}
\end{figure*}
\begin{figure}[!t]
\centering
\includegraphics[width=3.5in]{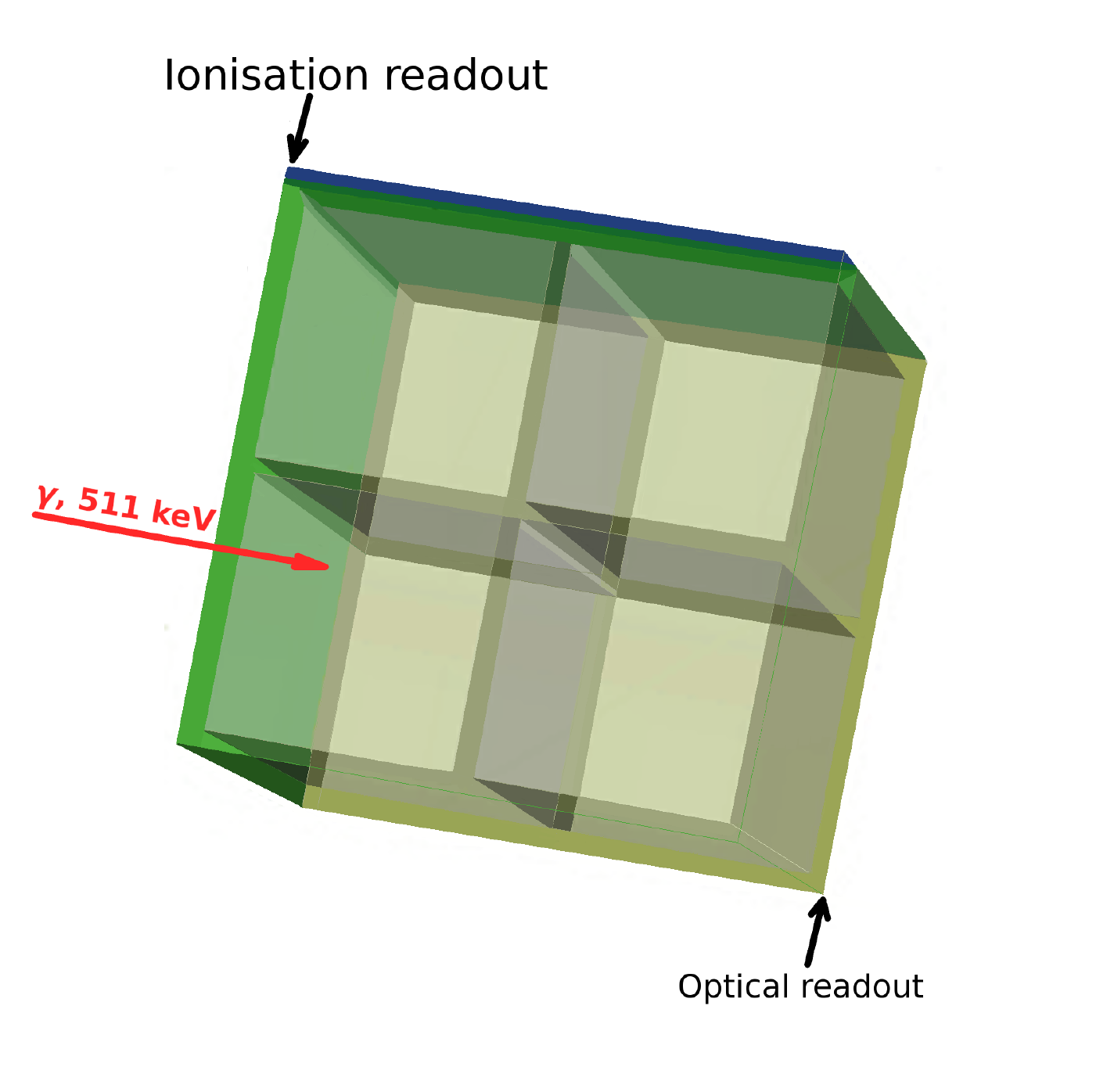}
\caption{Geometry of an elementary module used in the simulation for CaLIPSO PET. The module body is made from the alumina ceramic, sealed with the sapphire window from one side and metallized ceramic plate for the ionization readout from other side. The volume is filled with TMBi liquid and separated in four square cells by the white ceramic light guides.}
\label{fig3}
\end{figure}

The Monte Carlo simulation of the full size PET scanner was performed with the GATE 7.1 software~\cite{gate,gate:2010}, 
We have chosen cubic shape~(Fig.~\ref{fig2}) as it is possible thanks to the accurate reconstruction of the DOI inside a scanner cell, thus, exclusion of parallax effects. The main motivations for a cube shape of the scanner are the minimization of dead zones and the manufacture simplification. Moreover, study in~\cite{cube} has shown that the rectangular system with DOI capability has a higher signal-to-noise ratio for detection tasks and a lower bias at a given noise level for quantitation tasks than the same system without DOI.
 The CaLIPSO simulated scanner consists of 4 sectors by 5$\times$6 elementary modules~(Fig~\ref{fig3}). This determines the axial, 354~mm, and ``radial"\footnote{defined by the side of the cube}, 307~mm, sizes of the imager field of view~(FOV). 
In the current mechanical design the elementary detection module is made from the alumina ceramic body filled with the liquid TMBi. The volume is sealed with the transparent sapphire window using the metallic tightening from the external side. The MCP-PMT is coupled to the sapphire window with an optical gel. 
The read-out ionization pad structure is fabricated on the ceramic substrate and used to seal the volume from the inner side of the scanner. The elementary module is separated in four equal cells by 87\% reflecting ceramic light guides (thickness 1~mm) that focus the Cherenkov light on the photomultiplier surface. The module size, 59$\times$59~mm, is defined by the size of the selected PLANACON$^{TM}$ MCP-PMT by Photonis~\cite{Photonis}, The active part of MCP-PMT is 53$\times$53~mm, thus, dead zones are about 20\%. 
Laboratory test demonstrated that it is possible to reach 90\% efficiency for the 511~keV photons converted by the photo-ionization effects in TMBi~\cite{Ramos}, The module depth, 50~mm, allows to obtain a total detection efficiency more than 40\%.

\begin{figure}[!t]
\centering
\includegraphics[width=3.5in]{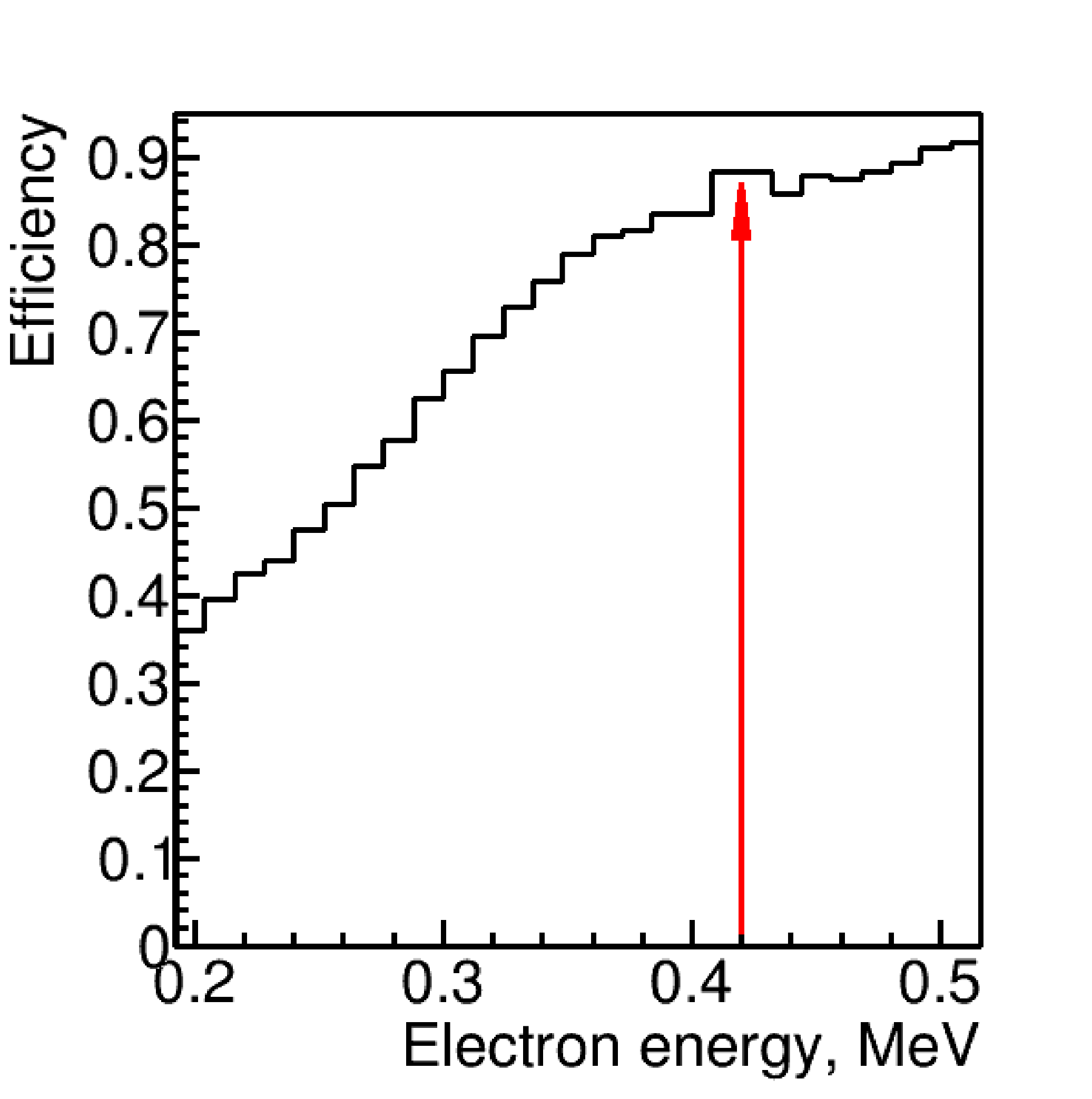}
\caption{Optical signal detection efficiency. Optical signal detection efficiency as a function of primary electron energy for the CaLIPSO PET scanner from Geant4 simulations. The red arrow corresponds to the most probable electron energy from 511~keV gamma photoelectric conversion. }
\label{fig4}
\end{figure}

The GATE digital detection model uses a dedicated parametrized description of the detector response for the ionization and light signal readout. These semi-analytic models are calibrated using detector prototypes when they are available and detailed Geant4 simulations in order to apply the following detector response functions:
\begin{itemize}
\item the efficiency of the optical signal detection as a function of energy obtained in Geant4 prototype simulations and published earlier~\cite{Ramos} illustrated in Fig~\ref{fig4};
\item 10\% (FWHM)~\cite{CaLIPSO} Gaussian blurring for the measured energy;
\item 1~mm (FWHM)~\cite{CaLIPSO} Gaussian blurring for the spatial coordinates.
\end{itemize} 

Coincidence events are selected within a 3~ns time window keeping multiple pairs of signles registered in the same time window if they are not detected in the same sector. The dead time is not yet taken into account but we expect that it would not significantly change the results. The first estimation shows that the non-paralyzable dead time $\tau_{dead}\sim$5~$\mathrm{\mu}$s, corresponding to drift time and not the pixelated detector dead time. The corresponding saturation rate is equal to $1/\tau_{dead}\sim$200~kHz per $1/4$ of scanner cell. Thus, saturation occupancy for the full scanner would be $480\times200$kHz$\sim$96~MHz while typical rate for a brain scan is at the level of 1~MHz. We also do not yet take into account the TOF potential of the CaLIPSO PET scanner but we discuss its influence on the Noise Equivalent Count Rate estimations later. This important advantage property of CaLIPSO PET scanner will be added to the simulations after conclusion on detector cell design.

In order to reduce the computation time, instead of positron source generation we simulate two 511~keV $\gamma$ photons emitted in opposite directions. In tests where it is relevant we take into account:
\begin{itemize}
\item \textit{acollinearity angle} introduced as a randomly chosen angle correction from Gaussian distribution with $\sigma=0.5$~degrees; 
\item \textit{positron range} set as a blurring of the source position according to distribution and values from~\cite{PositronRange},
\end{itemize}

\subsection{Image reconstruction}
The simulated data are reconstructed using the Customizable and
Advanced Software for Tomographic Reconstruction~(CASToR)~\cite{castor}, The CaLIPSO PET scanner is expected to have a spatial resolution of 1~mm$^3$, thus, a pseudo detection element would have the same size. This means that the full volume of TMBi liquid of CaLIPSO scanner contains $1.6\times10^7$ such elements. Thus, the number of lines of response~(LOR) is $\sim2.6\times10^{14}$, which is 5 orders of magnitude higher than for HRRT scanner. This amount of LORs makes sinogram based image reconstruction impossible. Thus, the list-mode OSEM algorithm is used along with a simple line projector~\cite{siddon} without including any resolution modeling. 
We perform 10 iterations with 16 subsets. The reconstructed images have a pixel size of 0.25~mm$\times$0.25~mm$\times$1~mm. In order to reduce voxel fluctuations we apply Gaussian smoothing with $\sigma=0.5$~mm on final brain images. This smoothing is not applied in the spatial resolution tests.

The first steps in the reconstruction are the scanner normalisation calculation and the sensitivity map creation. This stage remains difficult again due to the large data quantity. For example, to calculate the full scanner image normalization we need to generate $\sim2.6\times10^{16}$ events. Therefore, in this paper we use a simplified approach for the proof of concept and to illustrate the potential image spatial resolution: generate and reconstruct only one 2D axial slice. This approach allows to characterize image resolution specificities before having the complete and dedicated CaLIPSO image reconstruction package. For our first results we do not simulate and take into account the attenuation medium and we do not use the TOF information. Finally, only true coincidences are reconstructed as the reconstructed images are used for potential high resolution illustration.

\subsection{Performance tests}

In our case, the main parameters of interest are the Noise Equivalent Count Rates~(NECR) and image resolution. 

\subsubsection{Noise Equivalent Count Rates}
\label{sec:NECR}
In order to have an estimate of the expected image signal to noise ratio we use the following calculations of NECR parameter:
\begin{eqnarray}
NECR=\frac{T^2}{T+S+2R},
\label{eq:NECR}
\end{eqnarray}
where $T$, $S$ and $R$ are numbers of true, scatter and random coincidences respectively. The factor 2 corresponds to the on-line subtraction method for randoms correction, which is used in~\cite{HRRT}, reference in our comparisons. For consistency, we apply the same factor in the calculations. For the same reason, we also apply the same energy cut above 350 keV for coincidences selection. 

The simulation is performed with a cylindrical phantom NEMA-1994~\cite{NEMA:1994} with diameter and height of 200~mm. The phantom placed in the center of the FOV is filled with water and uniformly distributed radioactivity. The scatter, random, true fractions and NECR values are obtained for different activity concentrations in a range from 0.6~kBq/cc to 50~kBq/cc. 

The signal to noise ratio drops when spatial resolution increases as image voxel size becomes smaller and voxel variability increases. However, this ratio can be enhanced with time of flight method. To estimate TOF gain it is possible to use the the first order approximation~\cite{Conti2006}:
\begin{eqnarray}
NECR_{TOF}=\frac{D}{\Delta x}NECR,
\label{eq:NECRTOF}
\end{eqnarray}
where $D$ is the size of phantom and $\Delta x$ is is the localization uncertainty $\Delta x=c\Delta t/2$ related to the time resolution $\Delta t$ and the speed of light $c$. However, from the various studies estimating the TOF gain in terms of signal to noise ratio~(SNR)~\cite{Tomitani, Conti2006, Conti2013, Conti2015} depends on the way TOF is implemented in the reconstruction, the way random and scattered coincidences are handled and the actual count rates. We estimate the range of the NECR gain basing on approaches from these papers.

\subsubsection{Spatial resolution}
\label{sec:reso}
For the spatial resolution tests the full process of image reconstruction from the simulated data is applied. We perform three tests with [$^{18}$F] tracer:
\begin{enumerate}
\item The image resolution from a sample with $\sim8\cdot10^{6}$ detected coincidences we simulate several capillary sources of 0.2 mm diameter placed at different radial positions of the FOV of the CaLIPSO scanner.
\item To find the separation power between two capillary sources of 0.2 mm diameter we simulate them close to each other, at 1~mm and 2~mm distances between centers of the sources. We reconstruct $\sim5.5\cdot10^{6}$ coincidences.
\item A qualitative test is done with the one slice from mini Derenzo phantom~\cite{Derenzo} placed in a middle of FOV using $\sim1.5\cdot10^{8}$ detected coincidences. The phantom is a plastic cylinder of 45~mm diameter with hot rods of different diameters: 1.2~mm, 1.6~mm, 2.4~mm, 2.8~mm, 4.0~mm and 4.8~mm 
 The distances between centers of the rods are two times their diameters. 
\end{enumerate}
These tests aim to illustrate the high spatial resolution of the scanner. Thus, we reconstruct images containing true coincidences without quantification scaling. 

\subsection{Examples of human brain scans}
\label{sec:brain}
The simulations for human brain scans are done using one slice of the brain Zubal phantom~\cite{zubal} with a voxel size of 1.1~mm$\times$1.1~mm$\times$1.4~mm and with 18 different segmented brain structures. The activity distributions for two different tracers, [$^{11}$C]-PE2I 
 and [$^{18}$F]-FDG 
are defined from activity measurements from real HRRT scans. 

The aim of these tests is to illustrate the high image spatial resolution of the foreseen CaLIPSO PET scanner. Thus, we reconstruct high statistics images, i.e. containing $1.5\cdot10^{8}-1.6\cdot10^{8}$ true coincidences, and no quantification scaling is performed.

\section{Results}
\subsection{Noise Equivalent Count Rates}
\begin{figure*}[t]
\centering
\includegraphics[width=\textwidth]{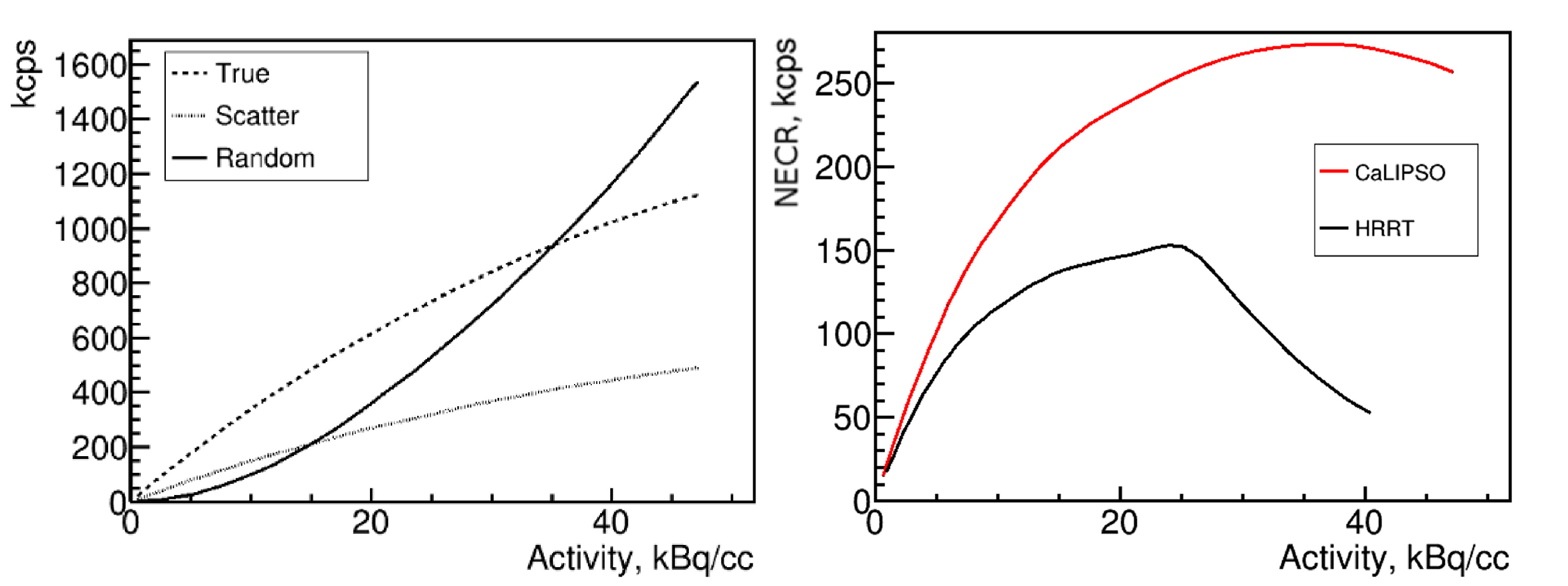}
\caption{Coincidence count rate of CaLIPSO. Simulated distributions as a function of activity for true, random and scatter coincidences (left) and Noise Equivalent Count Rates (right) for the simulated CaLIPSO scanner and the real HRRT scanner (Black curve in right figure is redrawn from~\cite{HRRT} by extracting data points with WebPlotDigitizer application~\cite{WebPlotDigitizer}).}
\label{fig5}
\end{figure*}

Fig~\ref{fig5},~left shows the curves for true, random and scatter coincidences as functions of activity and Fig~\ref{fig5},~right corresponds to the NECR curves according to Eq~\ref{eq:NECR} in ``Materials and Methods" section. One can see that the NECR values for the same activity are higher for the CaLIPSO system than for the HRRT. Thus, we expect to have an enhanced image contrast with the proposed CaLIPSO imaging system for the same voxel size. 
Additional increase of signal to noise ratio would be obtained with TOF technique. From Eq~\ref{eq:NECRTOF} a conservative estimate of $\Delta t=150$~ps gives $\Delta x=2.25$~cm and phantom size is $D$~=~20~cm. 
Thus, from the various studies estimating the TOF gain in terms of SNR~\cite{Tomitani, Conti2006, Conti2013, Conti2015}, it is expected to be in the range 5 to 10. This represents a significant increase of the signal to noise ratio.

\subsection{Image resolution}

\subsubsection{Test with capillary sources.}
\begin{figure}[!t]
\centering
\includegraphics[width=3.5in]{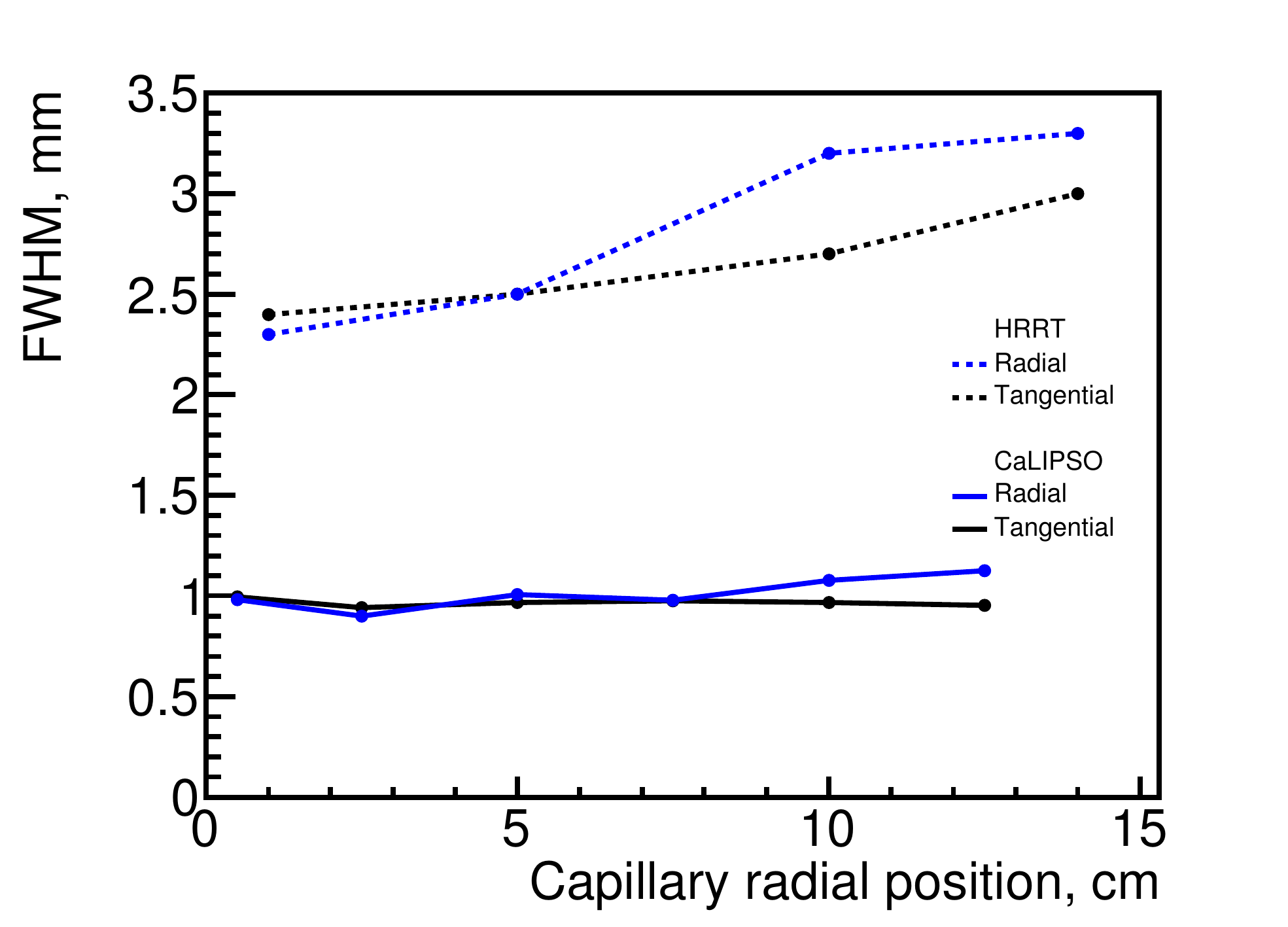}
\caption{Image resolution. Image resolution for the foreseen CaLIPSO scanner and its comparison to measurements from the HRRT scanner (Dashed curves are redrawn from~\cite{HRRT} by extracting data points with WebPlotDigitizer application~\cite{WebPlotDigitizer}).}
\label{fig6}
\end{figure}

The results obtained as explained in ``Materials and Methods" section for image resolution of capillary sources in radial and tangential directions are presented in Fig~\ref{fig6}, We obtain image resolution of $\sim$1~mm~(FWHM) with a slight degradation~($\sim$10\% relative error) toward peripheral FOV regions. In Fig~\ref{fig6} we also compare the results with HRRT system~\cite{HRRT} performance. One can observe that besides a lower image resolution the degradation toward peripheral FOV region is much higher for HRRT scanner.
\begin{figure*}[!t]
\centering
\includegraphics[width=\textwidth]{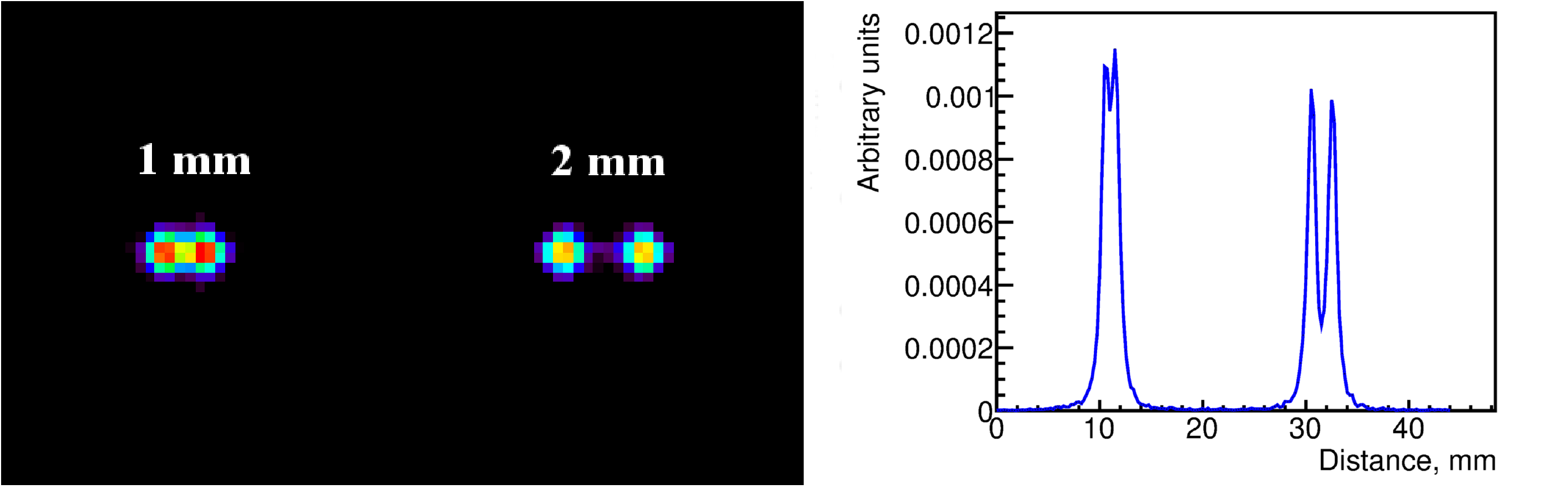}
\caption{Image resolution with two nearby capillaries. Two capillary sources generated at 1~mm and 2~mm distances between centres. Left figure corresponds to the reconstructed image, and right figure to a horizontal profile illustrating the separation of the peaks. No smoothing is applied on the reconstructed image.}
\label{fig7}
\end{figure*}

\subsubsection{Test with closely placed capillary sources}
The results for the test proposed in ``Materials and Methods" section with two capillaries placed closely one to each other are presented as a reconstructed image in Fig~\ref{fig7},~left and as a 1D profile in Fig~\ref{fig7},~right. One can notice that the two sources at 2~mm distance can be easily differentiated with a peak-valley ratio of $\sim$3 and a separation of $\sim$5~standard deviations. For the distance of 1~mm the difference between the peaks and the valley is about $\sim$1.2 and the separation is $\sim$2.4~standard deviations. 

\subsubsection{Test with the Derenzo phantom}
\begin{figure*}[!t]
\centering
\includegraphics[width=\textwidth]{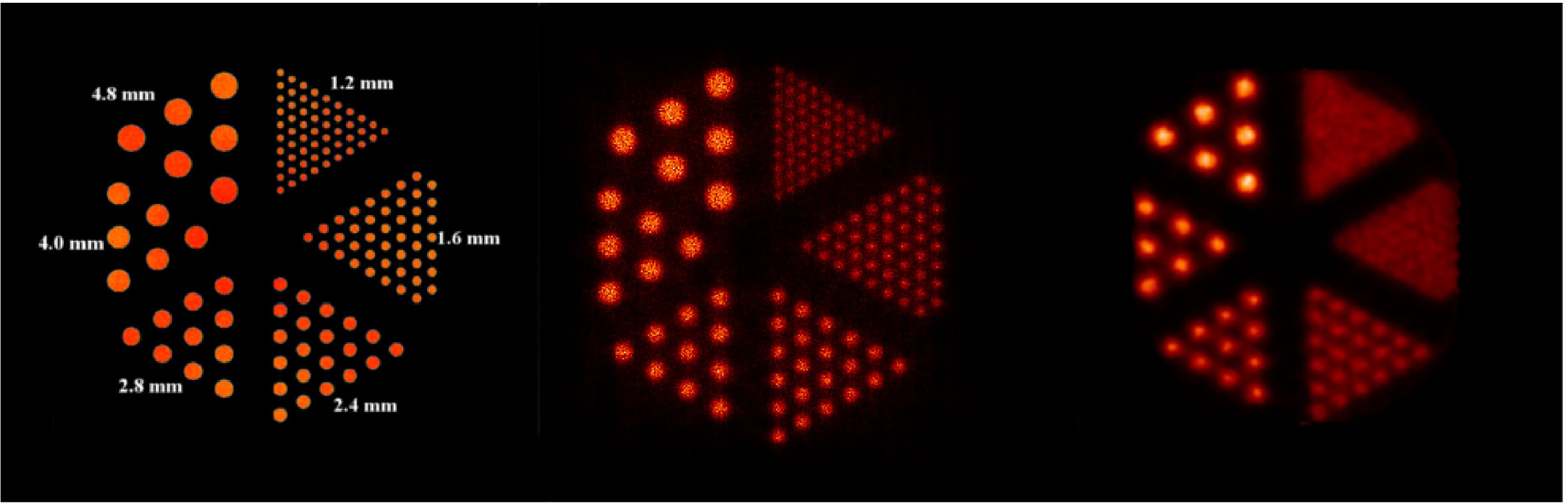}
\caption{Image resolution illustrated on Derenzo phantom. Simulated Derenzo phantom~(left) and the CaLIPSO reconstruction~(middle) and real measurement acquisition with the HRRT~(right). No smoothing is applied on the simulated images}
\label{fig8}
\end{figure*}

The generated distribution is illustrated in Fig~\ref{fig8},~left, while in the right figure we show the reconstructed image. One can observe that the smallest rods of 1.2~mm diameter can be clearly separated (Fig~\ref{fig8},~middle). 
Fig~\ref{fig8},~right shows the HRRT image obtained using real measurements.

\subsection{Examples of human brain scans}
\begin{figure*}[!t]
\centering
\includegraphics[width=\textwidth]{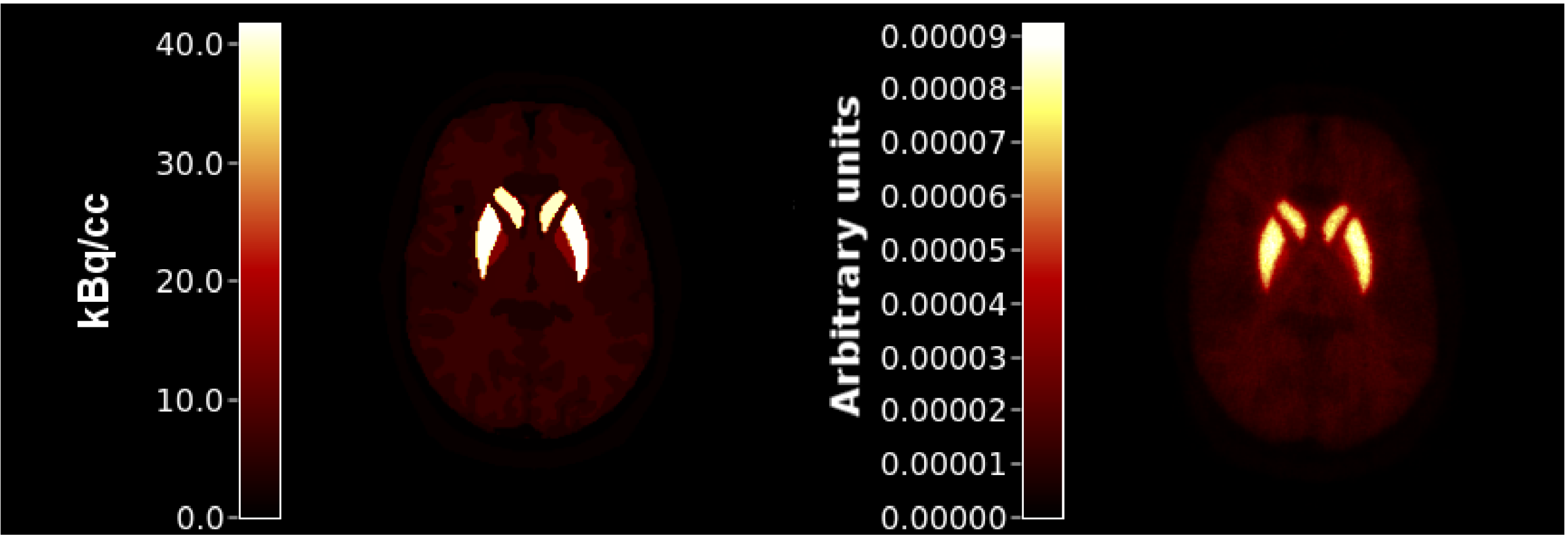}
\caption{Image resolution illustrated on Zubal phantom with [$^{11}$C]-PE2I tracer. Simulated (left) and CaLIPSO reconstructed (right) distributions of [$^{11}$C]-PE2I. Gaussian smoothing with $\sigma=0.5$~mm is applied to the reconstructed image.}
\label{fig9}
\end{figure*}

\begin{figure*}[!t]
\centering
\includegraphics[width=\textwidth]{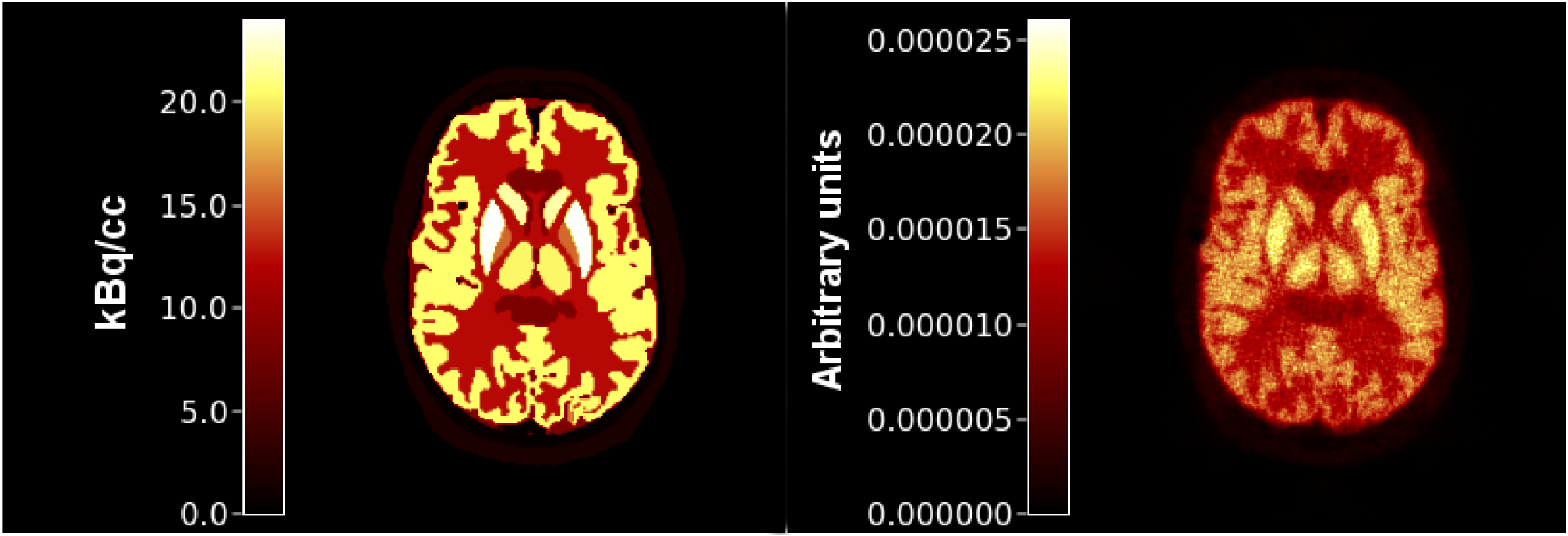}
\caption{Image resolution illustrated on Zubal phantom with [$^{18}$F]-FDG tracer. Simulated (left) and CaLIPSO reconstructed (right) distributions of [$^{18}$F]-FDG. Gaussian smoothing with $\sigma=0.5$~mm is applied to the reconstructed image.}
\label{fig10}
\end{figure*}

The brain simulations for [$^{11}$C]-PE2I and [$^{18}$F]-FDG tracers are presented in Fig~\ref{fig9} and \ref{fig10} respectively. Left figures correspond to the simulated activity distributions and right ones to the reconstructed images.
 
For the [$^{11}$C]-PE2I tracer (Fig~\ref{fig9}) one should keep in mind that the positron range of $^{11}$C is about 1~mm~(FWHM), which is compatible with the scanner spatial resolution. This means that for such type of tracers the images can not be refined further by improving the machine spatial resolution.

For the [$^{18}$F]-FDG tracer the situation is different. The positron range of $^{18}$F is $\sim$0.6~mm~(FWHM). Therefore, the high image resolution of the foreseen CaLIPSO scanner plays a key role, enabling a better separation of the different brain structures.

In Fig~\ref{fig9} and~\ref{fig10} the TOF techniques for image reconstruction are not taken into account and, thus, further improvement of image quality due to noise reduction would be possible. 
However, the simulation and reconstruction are performed without taking into account the attenuation, scatter and randoms contributions which would degrade the quality of the images. Nonetheless, from the results from NECR we expect that it would not be an issue.

\section{Discussion}

In this article we present the simulation study of the performance for the foreseen CaLIPSO scanner. The results exhibit the high potential of such a system. However, the studies with reconstructed images are restricted to one 2D slice due to the large dimensionality of the problem. 
It is worth noting that this problem is becoming more recurrent for new PET scanners with a number of detecting elements few orders of magnitude higher than in state-of-the art clinical scanners (either for very high spatial resolution or for extended axial FOV). We believe that it should not be an issue with the development of appropriate methodologies (beyond the scope of this paper) and increasing computer performances.


Several other parameters let aside in this paper should be taken into account in future studies, such as dead time of detection modules, TOF information, attenuation and scatter corrections of images and quantification scaling including normalisation. 

Another point to discuss is the complexity of the detection system and technology. The optical detection prototype is now operative~\cite{Ramos}, Currently we work on its improved version. The ionization detection prototype is under development and there are still several technological problems to be solved, including the ultrapurification of TMBi liquid and detection of low amplitude signals.

\section{Conclusion}

In this paper we present simulation results of the foreseen CaLIPSO PET scanner. The simultaneous detection of light and charge signals in CaLIPSO detectors leads to promising performances for a PET imager. Thanks to this detection technology we expect an image resolution of 1~mm~(FWHM) in the 3 dimensions as demonstrated in this work by simulated tests with capillary sources and the Derenzo phantom. The obtained NECR curve shows that the expected image signal to noise ratio is almost twice better than for the HRRT. This can be further increased by a factor 5-10 thanks to a TOF capability with a coincidence resolving time of 150~ps~(FWHM). 

Finally, reconstructed images of [$^{11}$C]-PE2I and [$^{18}$F]-FDG brain scans confirm the potential of the high spatial resolution expected with the CaLIPSO PET scanner.

\IEEEpubidadjcol

\section*{Acknowledgment}
This work was supported in part by the "Programme interdisciplinaire Technologie pour la Sante” of Commissariat \'{a} l’Energie Atomique et aux Energies Alternatives. Also authors would thank the support of The Neuropoole de Recherche Francilien and Laboratoire d'excellence Physique des deux infinis et des Origines.

\end{document}